\documentclass{llncs}

\usepackage[title,toc,titletoc,header]{appendix}
\usepackage{mathptmx}
\usepackage[utf8]{inputenc}
\usepackage{tikz}
\usepackage{listings}
\usepackage{url}
\usepackage{longtable}
\usepackage[english]{babel}
\usepackage[autostyle=true]{csquotes}
\usepackage{amssymb}
\usepackage{mathtools}
\usepackage{pgfplots}
\usepackage{todonotes}
\usepackage{wrapfig}
\usepackage{siunitx}
\usepackage[misc]{ifsym}

\pgfplotsset{
   compat=1.13, 
   /pgf/number format/.cd,
      fixed,
      set thousands separator = \thinspace, 
      min exponent for 1000 sep=4 
}
\usetikzlibrary{positioning}
\usetikzlibrary{arrows}
\usetikzlibrary{arrows.meta}

\newcommand{\plspec}{\textit{plspec}}
\newcommand{\prob}{\textsc{ProB}}
\newcommand{\swi}{\textit{SWI Prolog}}
\newcommand{\sic}{\textit{SICStus Prolog}}

\newcommand{\ourtool}{\textit{plstatic}}

\newcommand{\prespec}{precondition}

\newcommand{\prespecs}{preconditions}
\newcommand{\Prespecs}{Preconditions}
\newcommand{\postspec}{postcondition}
\newcommand{\postspecs}{postconditions}
\newcommand{\Postspecs}{Postconditions}
\newcommand{\prepostspecs}{pre- and postconditions}

\newcommand{\ignore}[1]{} 

\usepackage{multirow}
\usepackage{hyperref}
\usepackage[capitalise]{cleveref}

\usepackage{xcolor}
\definecolor{orcidlogocol}{HTML}{A6CE39}
\usetikzlibrary{svg.path}
\tikzset{
  orcidlogo/.pic={
    \fill[orcidlogocol] svg{M256,128c0,70.7-57.3,128-128,128C57.3,256,0,198.7,0,128C0,57.3,57.3,0,128,0C198.7,0,256,57.3,256,128z};
    \fill[white] svg{M86.3,186.2H70.9V79.1h15.4v48.4V186.2z}
                 svg{M108.9,79.1h41.6c39.6,0,57,28.3,57,53.6c0,27.5-21.5,53.6-56.8,53.6h-41.8V79.1z M124.3,172.4h24.5c34.9,0,42.9-26.5,42.9-39.7c0-21.5-13.7-39.7-43.7-39.7h-23.7V172.4z}
                 svg{M88.7,56.8c0,5.5-4.5,10.1-10.1,10.1c-5.6,0-10.1-4.6-10.1-10.1c0-5.6,4.5-10.1,10.1-10.1C84.2,46.7,88.7,51.3,88.7,56.8z};
  }
}
\renewcommand{\orcidID}[1]{%
  \resizebox{8px}{8px}{
      \href{https://orcid.org/#1}{\tikz[yscale=-1,transform shape]{\pic{orcidlogo}}}}%
}

\begin{document}
\title{Effectiveness of Annotation-Based Static Type Inference}
\author{Isabel Wingen, Philipp K\"orner $^\textrm{\Letter}$ \orcidID{0000-0001-7256-9560}}
\institute{
 Institut f\"{u}r Informatik, Universit\"{a}t D\"{u}sseldorf\\
  Universit\"{a}tsstr. 1, D-40225 D\"{u}sseldorf, Germany\\
 {\tt\scriptsize \{isabel.wingen,p.koerner\}@uni-duesseldorf.de}}

\maketitle

\begin{abstract}
Benefits of static type systems are well-known:
they offer guarantees that no type error will occur during runtime
and, inherently, inferred types serve as documentation on how functions are called.
On the other hand, many type systems have to limit expressiveness of the language
because, in general, it is undecidable whether a given program is correct regarding types.
Another concern that was not addressed so far is that,
for logic programming languages such as Prolog,
it is impossible to distinguish between intended and unintended
failure and, worse, intended and unintended success without additional annotations.

In this paper, we elaborate on and discuss the aforementioned issues.
As an alternative, we present a static type analysis which is based on \plspec{}.
Instead of ensuring full type-safety,
we aim to statically identify type errors on a best-effort basis
without limiting the expressiveness of Prolog programs.
Finally, we evaluate our approach
on real-world code featured in the SWI community packages
and a large project implementing a model checker.
\end{abstract}


\section{Introduction}%
\label{sec:motivation}

Dynamic type systems often enable type errors during development.
Generally, this is not too much of an issue
as errors usually get caught early by test cases or REPL-driven development.
Prolog programs however do not follow patterns
prevalent in other programming paradigms.
Exceptions are thrown rarely and
execution is resumed at some prior point via backtracking instead,
before queries ultimately fail (or succeed due to the wrong reason).
This renders it cumbersome to identify type errors,
their location and when they occur.

There has been broad research on type systems offering
a guarantee about the absence of type errors (briefly discussed in \cref{sec:position}).
Yet, in dynamic programming languages such as Prolog,
a complete well-typing of arbitrary programs is undecidable~\cite{pfenning1993undecidability}.
Thus, in order for the type system to work, the expressiveness of the language often is limited.
This hinders adaptation to existing code severely, and, as a consequence,
type errors are often ignored in larger projects.

At DECLARE'17, we presented \plspec{}~\cite{plspec}, a type system that uses
annotations in order to insert run-time type checks (cf. \cref{sec:plspec}).
During discussions, the point was raised that some type checks
could be made statically even with optional types.
This paper thus contributes the following:
\begin{itemize}
    \item A type analysis tool usable for \emph{any} unmodified Prolog program.
        It handles a proper \enquote{any} type and is extensible for any Prolog dialect (\cref{sec:analyser}).
    \item An empirical evaluation of the amount of inferred types using this tool (\cref{sec:evaluation}).
    \item Automatic inference and generation of \prepostspecs{} of \plspec{}.
\end{itemize}


\section{A Note on Type Systems and Related Work}%
\label{sec:position}

Static type systems have a huge success story,
mostly in functional programming languages like Haskell~\cite{jones2003haskell},
but also in some Prolog derivatives, such as Mercury~\cite{mercurytypes},
which uses type and mode information in order to achieve major performance boosts.
Even similar dynamic languages such as Erlang include a type specification language~\cite{Jimenez:2007:LST:1292520.1292523}.
Many static type systems for logic programming languages have been presented~\cite{pfenning1992types},
including the seminal works of Mycroft and O'Keefe~\cite{mycroft1984polymorphic},
which also influenced Typed Prolog~\cite{lakshman1991typed},
and a pluggable type system for Yap and SWI-Prolog~\cite{schrijvers2008towards}.

All type systems have some common foundations,
yet usually vary in expressiveness.
Some type systems \emph{suggest} type annotations
for functions or predicates,
some \emph{require} annotations of all predicates
or those of which the type cannot be inferred automatically
to a satisfactory level.
Yet, type checking of logic programs is, in general,
undecidable~\cite{pfenning1993undecidability}.
This renders only three feasible ways to deal with typing:

\begin{enumerate}
    \item Allow only a subset of types, for which typing is decidable, e.g., regular types~\cite{gallagher2004abstract}
        or even only mode annotations~\cite{rohwedder1996mode}.
    \item Require annotations where typing is not decidable without additional information.
    \item Work on a best-effort basis which may let some type errors slip through.
\end{enumerate}

Most type systems fall into the first or the second category.
Yet, this usually limits how programs can be written:
some efficient or idiomatic patterns may be rejected by the type system.
As an example, most implementations of the Hindley-Milner type system~\cite{milner1978theory}
do not allow heterogeneous lists, though always results in a well-typing of the program.
Additionally, most type systems refuse to handle a proper
\enquote{any} type, where not enough information is available
and arguments may, statically, be any arbitrary value.
Such restrictions render adaptation of type systems
to existing projects infeasible.
Annotations, however, can be used to guide type systems and allow more precise typing.
The trade-off is code overhead introduced by the annotations themselves,
which are often cumbersome to write and to maintain.

Into the last category falls the work of Schrijvers et al.~\cite{schrijvers2008towards},
and, more well-known, the seminal work of
Ciao Prolog~\cite{hermenegildo2012overview}
featuring a rich assertion language which
can be used to describe types.
Unfortunately, \cite{schrijvers2008towards}~seems to be abandoned after an early publication
and the official release was removed.
Ciao's approach, on the other hand, is very powerful,
but suffers due to incompatibilities with other Prolog dialects.

We share the reasoning and philosophy behind Ciao stated in~\cite{hermenegildo2012overview}:
type systems for languages such as Prolog must be optional
in order retain usefulness, power and expressiveness of the language,
even if it comes at the cost that not all type errors can be detected.
Mycroft-O'Keefe identified two typical mistakes type systems uncover:
firstly, omitted cases and, secondly, transposed arguments.
We argue that omitted cases might as well be intended failure
and, as such, should not be covered by a type system at all.
Traditional type systems such as
the seminal work of Mycroft-O'Keefe~\cite{mycroft1984polymorphic}
often are not a good fit,
as typing in Prolog is a curious case:
due to backtracking and goal failure,
type errors may lead to behaviour that is valid, yet unintended.


\paragraph{Backtracking.}
Prolog predicates are allowed to offer multiple solutions
which is often referred to as non-determinism.
Once a goal fails, execution continues at the last choice point
where another solution might be possible.
Thus, if a predicate was called incorrectly,
the program might still continue because another solution
is found, e.g., based on other input.
Consider an error in a specialised algorithm:
if there is a choice point,
a solution might still be found if
another, slower, fall-back implementation is invoked
via backtracking.
Such errors could go unnoticed for a long time
as they cannot be uncovered by testing
if a correct solution is still found in a less efficient manner.

\paragraph{Goal Failure.}
Most ISO Prolog predicates raise an error
if they are called with incorrect types.
However, non-ISO predicates usually fail as
no solution is found because the input does not match with any clause.
E.g., consider a predicate as trivial as \verb|member|:

\vspace{-.5mm}
\begin{verbatim}
  member(H, [H|_]).     member(E, [_|T]) :- member(E, T).
\end{verbatim}

\vspace{-.5mm}
\noindent Querying \verb|member(1, [2,3,4])|
will fail because \emph{the first argument is not in the list},
which is the second argument.
We name this \emph{intended failure}.
Yet, if the second argument is not a list,
e.g., when called as \verb|member(1, 2)|,
it will fail because \emph{the second argument is not a list}.
We call this \emph{unintended failure},
as the predicate is called \emph{incorrectly}.
The story gets even worse:
additionally to failure cases, there can also be unintended \emph{success}.
Calling \verb+member(2, [1, 2|foo])+ is not intended to succeed,
as the second argument is not a list, yet the query returns successfully.
Distinguishing between intended and unintended behaviour is impossible
as they use the same signal, i.e. goal failure (or success).
We argue that the only proper behaviour would be to raise an error on unintended input instead
because this most likely is a programming error.



In this paper, we investigate the following questions:
Can we implement an optional type system that supports
\emph{any Prolog dialect}?
How well does such a type system perform
and is a subset of errors that are identified on
\emph{best-effort basis} sufficient?
We think that the most relevant class of errors
is that an argument is passed incorrectly,
i.e. the type is wrong.
Thus, an important question is how precise type inference
by such a type system could be.
If it works well enough, popular error classes such as
transposed arguments,
as described by~\cite{mycroft1984polymorphic},
can be identified in most cases.


\section{Foundation: \plspec{}}%
\label{sec:plspec}

\plspec{} is an ad-hoc type system that executes type checks at runtime
via co-routining.
With \plspec{}, it is possible to add two kinds of annotations.
The first kind of annotation allows introduction of new types.
\plspec{} offers three different ways for this.
For our type system, we currently focus only on the first one
and implement shipped special cases that fall under the third category,
i.e. tuples, lists and compound terms:

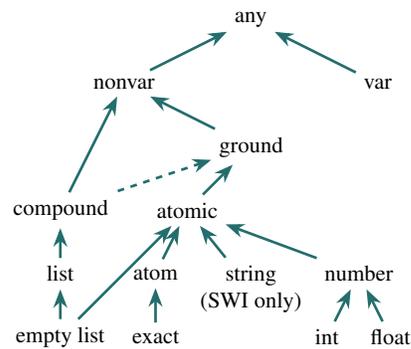
\begin{wrapfigure}[14]{R}{16em}
    \resizebox{16em}{!}{
\begin{tikzpicture}
\begin{scope}[every node/.style={}]
    \node (any) at (7,1.5) {any};
    \node (var) at (9,0.5) {var};
    \node (nonvar) at (5,0.5) {nonvar} ;
    \node (ground) at (7,-0.5) {ground} ;
    \node (compound) at (4,-1.5) {compound} ;
    \node (list) at (4,-2.5) {list} ;
    \node (emptylist) at (4,-3.5) {empty list} ;
    \node (atomic) at (6,-1.5) {atomic} ;
    \node (atom) at (5.5,-2.5) {atom} ;
    \node [align=center] (string) at (7,-2.74) {string\\(SWI only)} ;
    \node (exact) at (5.5,-3.5) {exact} ;
    \node (number) at (8.7,-2.5) {number} ;
    \node (int) at (8.2,-3.5) {int} ;
    \node (float) at (9.2,-3.5) {float} ;

\end{scope}

\begin{scope}[>={Stealth},
              every node/.style={fill=white,circle},
              every edge/.style={<-,draw={rgb:red,1;green,3;blue,3},very thick}]
     \path [<-] (any) edge (var);
     \path [<-] (any) edge (nonvar);
     \path [<-] (nonvar) edge (ground);
     \path [<-] (ground) edge (atomic);
     \path [<-] (atomic) edge (atom);
     \path [<-] (atomic) edge (number);
     \path [<-] (number) edge (int);
     \path [<-] (number) edge (float);
     \path [<-] (atom) edge (exact);
     \path [<-] (compound) edge (list);
     \path [<-] (nonvar) edge (compound);
     \path [<-] (list) edge (emptylist);
     \path [<-] (atomic) edge (emptylist);
     \path [<-] (atomic) edge (string);
     \path [<-,dashed] (ground) edge (compound);
\end{scope}
\end{tikzpicture}
}
\caption{Abstract Type Domain}%
\label{fig:domain}
\end{wrapfigure}

\begin{enumerate}
    \item recombination of existing types
    \item providing a predicate that acts as characteristic function
    \item rules to check part of a term and generate new specifications for sub-terms
\end{enumerate}

\plspec{}'s built-in types are shown in \cref{fig:domain}.
They correspond to Prolog types, with the addition of \enquote{exact},
which only allows a single specified atom (like a zero-arity compound),
and \enquote{any}, which allows any value.
Some types are polymorphic,
e.g. lists can be instantiated to lists of a specific type.
There are also two combinators, \verb|one_of|
that allows union types as well as \verb|and|,
which is the intersection of two types.

Combination of built-in types is certainly very expressive.
While such structures cannot be inferred easily without prior definition,
as a realistic example,
it is possible to define a tree of integer values
by using the \verb|one_of| combinator as follows:

\begin{verbatim}defspec(tree, one_of([int, compound(node(tree, int, tree))])).\end{verbatim}

Valid \verb|tree|s are \verb|1|, \verb|node(1, 2, 3)|,
\verb|node(node(0, 1, 2), 3, 4)| but not, e.g.
\verb|tree(1, 2, 3)|, where the functor does not match,
or \verb|node(a, b, c)| which stores atoms instead
of integer values.
Note that it is also possible to use a wildcard type to define
a tree \verb|tree(specvar(X))|,
which passes the variable down into its nodes. \verb|specvar|s are a placeholder
to express that two or more terms share a common, but arbitrary type.
This can be used to define template-like data structures
which can be instantiated as needed, e.g., as a \verb|tree(int)|.

The second kind of annotations specifies how predicates may be called
and, possibly, what parameters are return values.
We re-use two different annotations for that:

\begin{enumerate}
    \item \emph{Preconditions} specify types for all arguments of a predicate.
        For a call to be valid, at least one precondition has to be satisfied.
    \item \emph{Postconditions} add promises for a predicate:
        if the predicate was called with certain types and if the call was successful,
        specified type information holds on exit.
\end{enumerate}

Both pre- and postconditions must be valid for every clause of the specified predicate.
Consider a variation of \verb|member/2|, where the
second argument \textit{has to be} a list of atoms, and the first argument can
either be an \verb|atom| or \verb|var|:
\begin{verbatim}
atom_member(H,[H|_]).   atom_member(E,[_|T]) :- atom_member(E,T).
\end{verbatim}
Instead of checking the terms in the predicate,
type constraints describing intended input are added via \plspec{}'s pre- and postconditions.
The following
preconditions express the valid types one has to provide:
the first argument is either a variable or an atom,
and the second argument must be a list of atoms.
\begin{verbatim}
:- spec_pre(atom_member/2, [var, list(atom)]).
:- spec_pre(atom_member/2, [atom, list(atom)]).
\end{verbatim}

As the second argument is always a ground list of atoms, we can assure
callers of \verb|atom_member/2|, that the first term is bound after the
execution using a postcondition:

\begin{verbatim}
:- spec_post(atom_member/2, [var, list(atom)], [atom, list(atom)]).
\end{verbatim}

Postconditions for a predicate are defined using two argument lists:
they are read as an implication.
For \verb|atom_member/2| above, this means that
\enquote{if the first argument is a variable and the second argument is a list of atoms,
and if \texttt{atom\_member/2} succeeds, it is guaranteed that the second argument is still a list of atom,
but also that the first argument will be bound to an atom}.
If the premise of the postcondition does not hold or the predicate fails,
no information is gained.

\paragraph{Extensions to \plspec{}.}%
\label{plspec-extensions}


The traditional understanding if there are two instances of the same type variable,
e.g. in a call such as \verb|spec_pre(identity/2, [X, X])|,
is that both arguments \emph{share all types}.
Yet, we want to improve on the expressiveness of, say, \verb|spec_pre(member/2, [X, list(X)])|,
and allow heterogeneous lists.
This extension is not yet implemented in \plspec{} itself
and is only part of the static analysis in \ourtool{}.
In order to express how the type of type variables is defined,
we use \verb|compatible| for the homogeneous and \verb|union| for the
heterogeneous case.

If a list is assigned the type \verb|list(compatible(X))|, every item in the
list is assigned the type \verb|compatible(X)|. Now \ourtool{} checks whether all
these terms share all types, thus enforcing a homogeneous list.
If a list is assigned the type \verb|list(union(X))|, every item in
the list is assigned the type \verb|union(X)|. But instead of a type intersection,
\ourtool{} collects the types of these terms and builds a union type.

To give an example for the semantics of compatible and union,
the list \verb|[1, a]| has the \emph{inner} type \verb|one_of([int, atom])|
under the semantics of a union, and results in a type error (as the intersection of int and atom is empty)
if its elements should be compatible.
A correct annotation for \verb|member/2| would be the following \postspec{}:\\
\verb|spec_post(member/2,[any,list(any)],[compatible(X),list(union(X))])|, i.e.,
the list is heterogeneous, and the type of the
first argument must occur in this list.

\usetikzlibrary{arrows.meta}
\tikzset{%
  >={Latex[width=2mm,length=2mm]},
            base/.style = {rectangle, rounded corners, draw=black,
                           minimum width=2cm, minimum height=.8cm,
                           text centered, font=\sffamily},
          module/.style = {base, fill=blue!30},
             dir/.style = {base, fill=green!30},
           }

\section{Our Type System}%
\label{sec:analyser}

In the following, we describe a prototype named \ourtool{}.
It uses an abstract interpreter in order to collect type information on Prolog programs
and additionally to identify type errors on a best-effort (i.e., based on available type information due to annotations) basis,
without additional annotations.
The tool is available at \url{https://github.com/isabelwingen/prolog-analyzer}.
Due to page limitation, we can only present some points we deem important.

\paragraph{Purpose and Result.}
The tool \ourtool{} performs a type analysis on the provided code.
All inferred information can be written out
in form of annotations in \plspec{} syntax,
or HTML data that may serve, e.g., as documentation.
Naturally, \ourtool{} shows an overview of type errors, which were found
during the analysis.
\ourtool{} is not intended to uncover all possible type errors.
Instead, we are willing to trade some false negatives for the absence of
false positives, as they might overwhelm a developer in pure quantity.
Whether true programming errors can be discovered is discussed
in \cref{sec:evaluation}.

As typing can be seen as a special case of abstract interpretation~\cite{cousot1997types},
we use \plspec{}'s annotations to derive an abstract value,
i.e. a type, for terms in a Prolog clause.
Abstract types correspond to the types shown in \cref{fig:domain},
where a type has an edge pointing to a strict supertype.
However, as distinguishing ground from nonvar terms often is important,
compound terms are tried to be abstracted to the ground type first, represented by the dashed edge.
We use the least upper bound and greatest lower bound
operations as they are induced by the type subset relation.
This analysis is done statically and without concrete
interpretation of Prolog code,
based on \plspec{} annotations
and term literals.

\paragraph{Annotations.}


\ourtool{} works without additional annotations in the analysed code.
It derives type information from
(a large subset of) built-in (ISO) predicates, that
we manually provided \prepostspecs{} for.
We also annotated a few popular libraries, e.g. the lists library.
For predicates lacking annotations, types can be derived
if type information exists for predicates called in their body,
or can be inferred from unification with term structure in the code.
Derived types describe intended success for the unannotated predicate.
Naturally, precision of the type analysis improves with more annotations.

\subsection{Tool Architecture}

\ourtool{} is implemented in Clojure.
An alternative was to implement a meta-interpreter in Prolog.
A JVM-based language allows easier integration into text editors,
IDEs and potentially also web services.
However, this requires to extract a representation of the Prolog program.
We decided against parsing Prolog due to operator definitions and loss of term expansion\footnote{Term expansion is a mechanism that allows source-to-source transformation.}.
Instead, we add a term expander ourselves before we load the program.
It implements \plspec{}'s syntax for annotations
and extracts those alongside the program itself.
All gathered information is transformed to
edn\footnote{https://github.com/edn-format/edn}.

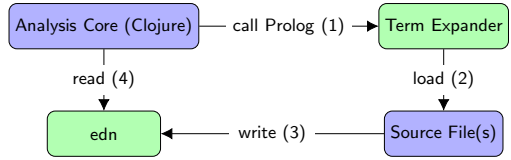
\begin{figure}[t]
\centering

\resizebox{0.55\textwidth}{!}{
\begin{tikzpicture}[node distance=1.0cm, every node/.style={fill=white, font=\sffamily}, align=center]
   \node (Clojure)    [module]                                          {Analysis Core (Clojure)};
   \node (edn)        [dir, below of=Clojure, xshift=0.0cm, yshift=-0.9cm]  {edn};
   \node (SourceDir)  [dir, right of=Clojure, xshift=5.0cm, yshift=0.0cm]  {Term Expander};
   \node (Prolog)     [module, below of=SourceDir, xshift=0.0cm, yshift=-0.9cm] {Source File(s)};
    \draw[->]             (Clojure) -- node[text width=2cm] {\mbox{call Prolog~(1)}} (SourceDir);
    \draw[->]             (Clojure) -- node[text width=2cm] {read~(4)} (edn);
    \draw[->]             (SourceDir) -- node[text width=2cm] {load~(2)} (Prolog);
    \draw[->]             (Prolog) -- node[text width=1.3cm] {write~(3)} (edn);

\end{tikzpicture}}
 \caption{Tool Architecture}
 \label{fig:flow}
 \end{figure}

\ourtool{} consists of two parts pictured in \cref{fig:flow}: a binary (jar) that contains the
static analysis core, and a term expander written in Prolog,
The analysis core is started with parameters specifying
the path to a Prolog source file or directory and a Prolog dialect (for now, ``swipl'' or ``sicstus'').
Additionally, the path to the term expander can be passed as an argument as well,
if another syntax for annotations than \plspec{}'s is desired.

Regarding module resolution, special care has to be taken when an entire directory is analysed:
when modules are included, it is often not obvious
where a predicate is located.
It can be hard to decide whether a predicate is user-defined,
shipped as part of a library or part of the built-in predicates
available in the user namespace.
Thus, when the edn-file is imported, a data structure is kept
in order to resolve calls correctly.


As our evaluation in~\cref{sec:evaluation} uses untrusted third-party code,
we take care that the Prolog code, that may immediately run when loaded, is not executed.
Instead, the term expander does not return any clause,
effectively removing the entire program during compilation.
Trusted term expanders can be loaded beforehand
if required.

\subsection{Analysis}
Our approach to type inference implements a classical abstract interpreter.
Each clause is analysed individually in a first phase.
We use \plspec{}'s annotations of the clause and the sub-goals to derive
an abstract type domain for all terms in the clause.
In a second phase, those results are combined:
After the first phase, we have obtained a typing for every clause, which
describes the types that the terms have after a
successful execution of the clause.
The inferred type information for all clauses of a predicate, can be stored as a \postspec{}.
This \postspec{} may be more accurate than the already provided one. In
this case, the analysis of a predicate would in turn improve the
analysis result for clauses that call that predicate.

For this reason, \ourtool{} works in two phases:
first, clause-local analysis that is based on already known information,
and, second, merging information of all clauses of a single predicate,
propagating newly gained information to the caller(s).
Without the presence of a \verb|one-of| combinator,
this would guarantee a fixed point as a result of the analysis.
As we cannot infer recursive datatypes yet,
which might result in infinite \verb|one-of|-sequences,
we limit the number of steps in order to ensure termination.


\paragraph{Example: Rate My Ship}
The following code will accompany us during this section.
\begin{lstlisting}
ship(Ship) :- member(Ship, [destiny, galactica, enterprise]).
rating(stars(Rate)) :- member(Rate, [1,2,3,4,5]).
rate_my_ship(S,R) :- ship(S), rating(R). \end{lstlisting}

\subsubsection{Preparation}

For every loaded predicate, we check, if there are \prepostspecs{} already specified,
ones provided by the user or our own manual annotations of ISO predicates.
Otherwise, they are created containing any-types during the preparation as follows:
all literals, e.g., lists, compound or atomic terms, in the clause head
are considered: their type is already known after loading the program.
For variable literals, however, we initially assume the type \verb|any|.
Additionally, if not annotated otherwise,
we assume that a clause may be called by a variable.
Based on this information, we create initial \prepostspecs{}
for all predicates, considering the \emph{entire} argument vector.




Below, we show the generated specs for our example after the preparation step:
\begin{lstlisting}
:- spec_pre(ship/1, [any]).
:- spec_post(ship/1, [any], [any]).
:- spec_pre(rating/1, [one_of([var, compound([stars(any)])])]).
:- spec_post(rating/1, [any], [compound([stars(any)])]).
:- spec_pre(rate_my_ship/2, [any, any]).
:- spec_post(rate_my_ship/2, [any, any], [any, any]).  \end{lstlisting}

\begin{figure}[t]
\centering
\resizebox{0.75\textwidth}{!}{
\begin{tikzpicture}[node distance=0.8cm, every node/.style={fill=white, font=\sffamily}, align=center]
   \node (L)    [module]  {brother(Lore,Data)\\ \{:dom tuple([atom,atom])\}};
   \node (H)    [module, xshift=-5cm, yshift=0.0cm]  {Lore\\ \{:dom atom\}};
   \node (T)    [module, xshift=5cm, yshift=0.0cm]   {Data\\ \{:dom atom\}};
   \draw[->]    (H) -- node[text width=1cm] {:arg 0} (L);
   \draw[->]    (T) -- node[text width=1cm] {:arg 1} (L);

\end{tikzpicture}}
    \caption{An Example Environment (Using edn-Formatted Maps)}
 \label{fig:env}
 \end{figure}
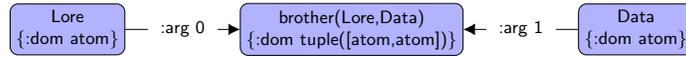

\subsubsection{Phase 1: Clause-Local Analysis}
\label{sec:algorithm}
Because of the nondeterministic nature of Prolog,
it is not sufficient to store the current type for a variable at a given point:
we also have to consider relationships between several terms that are caused by unification.
Such relationships are stored in an environment, for which we use a directed graph per clause.
The inferred types of the terms are stored in the vertices. 
Relationships between terms and sub-terms
e.g. \texttt{[H|T]}, where head and tail might have a dependency
on the entire list term (e.g., \verb|list(int)|),
or \postspecs{}
are saved as labelled edges between the term vertices.
An example showing 
the structure of a compound term \verb|brother(Lore, Data)| is given in~\cref{fig:env}.

During the analysis of a clause, the type domains of the
terms are updated and their precision is improved.
We assume that each predicate call in the body has to succeed,
and gather information from their \prepostspecs{}.
When new type information about a term is gained,
the greatest lower bound is calculated by
intersecting both domains.
When considering variables in Prolog however,
this comes with some pitfalls that are discussed in more details
in \emph{Step 2}.
If the type intersection is empty,
no concrete value is possible for the Prolog term
and a type error is reported.
However, this relies on the assumption that all given annotations are correct.

\paragraph{Step 1: Clause Head.}
The environment is initialised with all terms occurring in the head of the clause.
Information about the head of the clause can be derived
from the \prespecs{}.
According to \plspec{}, at least one \prespec{} must be fulfilled.

This raises the issue of tuple distributivity.
Consider 
a predicate \verb|cake(X, Y)| that is annotated with the \prespecs{}
\verb|[atom, int]| and \verb|[int, atom]|.
This means that \verb|cake/2| expects an atom and an integer, no matter the order.
For both \verb|X| and \verb|Y|, one could derive \verb|one_of([atom, int])| as type information.
However, this would render \verb|X=1,Y=2| to be valid input,
as the individual type constraint are fulfilled,
yet, the original \prespec{} is violated.

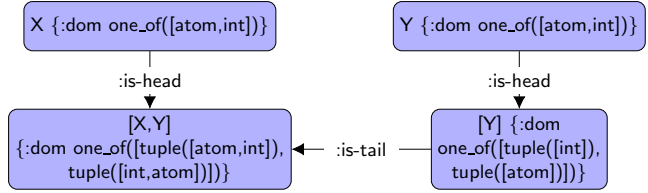
\begin{figure}[t]
\centering
    \resizebox{0.7\textwidth}{!}{
\begin{tikzpicture}[node distance=1.0cm, every node/.style={fill=white, font=\sffamily}, align=center]
   \node (X)       [module]  {X \{:dom one\_of([atom,int])\}};
   \node (Y)       [module, right of=X, xshift=5cm]  {Y \{:dom one\_of([atom,int])\}};
   \node (XY)      [module, below of=X, yshift=-1cm]  {[X,Y]\\ \{:dom one\_of([tuple([atom,int]),\\ tuple([int,atom])])\}};
   \node (LY)      [module, below of=Y, yshift=-1cm]  {[Y] \{:dom\\ one\_of([tuple([int]),\\ tuple([atom])])\}};
   \draw[->]    (X) -- node[text width=2cm] {:is-head} (XY);
   \draw[->]    (Y) -- node[text width=2cm] {:is-head} (LY);
   \draw[->]    (LY) -- node[text width=1cm] {:is-tail} (XY);

\end{tikzpicture}}
    \caption{Environment with a Watcher (Using edn-Formatted Maps)}
 \label{fig:watcher}
\end{figure}

As we aim at keeping the most precise type information possible,
we create an artificial tuple containing all arguments,
whose domain is a union-type containing all supplied \prespecs{}.
This artificial term functions as a \enquote{watcher},
and ensures all type constraints.
For the \verb|cake| predicate, the term \verb|[X,Y]| is added to the environment,
along with its type
\verb|one_of([tuple([atom,int]), tuple([int,atom])])|.
Once we know a more specific type for, e.g., \verb|Y|, we can derive which
option must be valid for the \enquote{watcher}, and
we can derive a type for \verb|X|.
The environment is pictured in \cref{fig:watcher}.

Due to page limitations, we only consider the environment of \verb|rate_my_ship/2| here:
in this step, it infers types for \verb|S|, \verb|R| and the entire argument vector \verb|[S,R]|.
%
%

\paragraph{Step 2: Evaluate Body.}
We analyse the body step by step, making use of (generated or annotated) pre- and postconditions of all encountered sub-goals.
This allows us to refine the type step by step: for example, if \verb|member(X,L)| is called,
one can infer that \verb|L| must be a list on success, even if no information on the variable was known before.
On the first occurrence of a term, it is added to the environment.
Similarly to the clause head, at least one \prespec{} of the sub-goal must be
compatible with the combination of the arguments it is called with.
Otherwise, for the example calling member, if \verb|L| is known not to be a list but, e.g., an integer,
a type error is raised.

The analysis does not step into the sub-goal,
and only uses \prepostspecs{}.
A \postspec{} specifies type constraints on a term
after the called predicate succeeds.
Thus, it is checked which premises of \postspecs{} are fulfilled.
Then, the greatest lower bound of the current type domain
and the possible conclusion of the \postspecs{} is calculated in order to improve precision.
An example is shown in \cref{tbl:envship}.

\begin{table}[t]
    \caption{Environment for \texttt{rate\_my\_ship/2}}
    \label{tbl:envship}
    \centering
\setlength{\tabcolsep}{4.5pt}
\begin{tabular}{l l l l}
    \hline\hline
    Variable Term & Clause Head & after 1st sub-goal & after 2nd sub-goal \\\hline
    [S, R]        & tuple([any, any]) & tuple([any, any]) & tuple([any, any]) \\ \relax
    [R]           & tuple([any]) & tuple([any]) & tuple([any]) \\
    R             & any & any & compound(star([any])) \\
    S             & any & any & any \\
    \hline\hline
\end{tabular}
\end{table}

%
%

\paragraph{Type Variables.}
We have introduced two new kinds of type variables (cf.~\cref{plspec-extensions}):
\verb|union| and \verb|compatible|.
It is possible to use \verb|union(X)| or \verb|compatible(X)|,
where \verb|X| is a type variable.
Both are placeholders for yet unknown types and express two different
relationships between terms:

Every term that is assigned the type \verb|union(X)| contributes to the
definition of the type that is \verb|X|.
The connection is made by adding a labelled edge \verb|:union| between the term and \verb|X|.
Then, the domain of all contributing terms is calculated as described.
At the end of the analysis step,
the union type of the variable \verb|X| 
is inferred via the least upper bound of all connected terms.
As an example, if an integer and an atom is part of the same union type,
it will result in \verb|one_of(int, atom)|.

On the other hand, terms that are assigned the type \verb|compatible(X)|, must
be compatible with all other terms that are assigned that type.
This implies that their intersection must not be empty.
As with the \verb|union| type, we create a labelled edge \verb|:compatible|
connecting the term to \verb|X|.
These edges are processed \emph{after} all union edges have been visited.
For example, if a known atomic value and a known integer have to be compatible within
the same type variable, we can infer that both values are integer,
as it is the intersection of both types.

In order to determine the type of a type variable,
it is required to know all contributing terms.
Thus, for compound or a list terms of a known size,
the assigned type is passed down to its sub-terms using
the mechanisms described above.
Yet, even if we know that \verb|L| is a list of \verb|union(X)|,
we do not know the list items yet --
even worse, the variable may only be bound later on!
This requires an additional step in order to ensure that the domain
for the type variable \verb|X| is compiled correctly:
we opted to add a \verb|:has-type| edge to the environment,
which connects a Prolog variable, e.g. \verb|T|,
to an artificially created variable \verb|T__<uuid>| storing the inner type,
i.e. \verb|union(X)| in the example above.
Whenever the domain of a connected variable is updated, so is the type variable itself.
Effectively, this delays the computation of the actual type variable.
The artificial list type variable then is connected with \verb|union(X)|.
For \verb|compound| and \verb|tuple| \emph{type specifications},
an artificial term is created and linked to the variable term
via a special edge.
This is required to mimic unification of Prolog variables.
Whenever the domain of the variable term is updated,
the artificial term's domain is updated as well.
Finally, the information is propagated into the corresponding sub-terms if required.

Have a look at \verb|member/2| used in the body of
\verb|ship/1|. The provided \postspec{}~is\\
\verb|post_spec(member/2, [any, any], [compatible(X), list(union(X))]).|\\
Therefore, after analysing the body of \verb|ship/1|, we know the following:

\begin{enumerate}
    \item The second argument of \verb|member| contributes to the variable \verb|X| in form of a \emph{union}.
        We learn that \verb|X| is either \verb|destiny|, \verb|galactica| or \verb|enterprise|.
    \item We learn that the variable \verb|Ship| must be compatible with \verb|X|,
        so it must be one of the three atoms named above.
\end{enumerate}



\paragraph{Step 3: Term Relationships.}
After analysing the body, all terms in the clause are included in the environment.
Then, nodes that may be destructured, i.e. lists and compound terms, are looked up in the graph.
As sub-terms, e.g. \verb|X| in \verb|a(X)|, can be used individually in subsequent sub-goals,
i.e. without the wrapping functor \verb|a(...)|,
inferred information has to be propagated back to the larger compound term.
We introduce the following edges in order to provide the necessary mechanism:

For lists, we extract the head and tail terms and add them to the environment,
if they are not already contained.
Those terms are marked with special edges \verb|:is-tail| and \verb|:is-head| (cf. \cref{fig:watcher})
pointing to the original list.
For compounds, we add the argument terms to the environment and store the
position of every term in the compound by adding an edge \verb|:pos| (cf. \cref{fig:env}).

For \verb|rate_my_ship/2|, three edges are added due to this step:
the environment already contains the argument vector \verb|[S,R]| after \emph{Step 1}.
We add that \verb|S| is the head item, that \verb|[R]| is the tail of the list,
and that \verb|R| is the head of the tail \verb|[R]|.

\paragraph{Prolog Variables.}
The any-type can be split into two disjoint sets:
variables and non-variable terms.
After processing a sub-goal, non-variable terms can only gain precision.
Variables, however, have the unique property that their type can change,
as they can be bound to, say, an atom, which is \emph{not} a sub-type.
To take this into account, a different intersection
mechanism is required for variables:

\begin{itemize}
  \item \Prespecs{} of the \emph{currently analysed} predicate may render a variable non-variable.
  \item \Prespecs{} of a \emph{called sub-goal} cannot render a variable term non-variable.
  \item \Postspecs{} of a \emph{called sub-goal} may render a variable term non-variable.
  \item Once a Prolog variable is bound to a non-variable, it behaves like any non-variable.
\end{itemize}

\paragraph{Step 4: Fixed-Point Algorithm.}
During the prior steps, we added edges to the environment. These are now used to update
the types of the linked terms.
If the environment no longer changes, we have consumed all collected
knowledge and have found a preliminary result for a clause.

For example, in \verb|rate_my_ship/2|, we will update the tuples
\verb|[R]| and \verb|[S,R]| once we learn that \verb|R|
must be of the form \verb|compound(stars([any]))|.

\subsubsection{Phase 2: Global Propagation of Type Information}
During the local analysis, each clause was inspected in isolation.
The type domains in the returned environments
contain the types after a successful execution of a clause \textit{with the knowledge gained so far}.
The gathered information then must be propagated to the caller of the corresponding predicate
in order to improve the precision of the type inference.

Each resulting environment can be used to generate the conclusion of a \postspec{}.
If a predicate succeeds, at least one of its clauses succeeded.
As \postspecs{} must be valid for the entire predicate,
the conclusion of a new \postspec{} is the union of 
all conclusions of the corresponding clauses.
This newly gained knowledge (in form of a \postspec{}) is added to the analysed data for every predicate.
Afterwards, both local analysis and global propagation are triggered,
until a fixed-point is reached.
Inferred \prepostspecs{} can be written out after analysis in \plspec{}'s syntax.

\paragraph{Example: append/2.}
Consider the append program:
\begin{verbatim} append([], Y, Y).    append([H|T], Y, [H|R]) :- append(T, Y, R).  \end{verbatim}
For the first clause, \ourtool{} would derive the types
\verb|[list(any), any, any]|.
For the second clause, we gain no additional information from the body, because
\verb|append/2| is calling itself, so we derive the types
\verb|[list(any), any, list(any)]|.
To create a conclusion of a \postspec{} for the predicate, we need to combine
the results of the two clauses. Unfortunately, as the type of the third argument
is \verb|any| in one case, it swallows the more precise type \verb|list(any)|.
We obtain the following conclusion:
\verb|[list(any), any, any]|.
While the \emph{intention} is that the second and third arguments are lists as well,
this cannot be inferred without annotations.

As you have probably noticed, \ourtool{} has not yet found the accurate
type \verb|atom| for \verb|S| or \verb|R| in \verb|rate_my_ship/2|.
This is because the \prepostspecs{} of \verb|ship/1| have not been updated yet,
so \ourtool{} has no way of knowing that \verb|S| is an atom.
In the first phase, we have concluded that the argument given to \verb|ship/1|
must be of type \verb|atom| after a successful execution.
% ...
As \verb|ship/1| has only one clause, we can infer the \postspec{}:
\verb|:- post_spec(ship/1, [any], [atom]).|
Analogously, we obtain
\verb|:- post_spec(rating/1, [any], [compound(stars([atom]))]).|

The propagation of the newly gained knowledge is shown in \cref{tbl:envship2}.
\begin{table}[t]
    \caption{Environment for \texttt{rate\_my\_ship/2}}
    \label{tbl:envship2}
    \centering
\setlength{\tabcolsep}{4.5pt}
\begin{tabular}{l l l l}
    \hline\hline
    Variable Term & Newly Gained Knowledge                & After Propagation \\ \hline
    [S, R]        & tuple([any, compound(star([any]))])   & tuple([\emph{atom}, compound(star([\emph{atom}]))]) \\ \relax
    [R]           & tuple([compound(star([any]))])        & tuple([compound(star([\emph{atom}]))]) \\
    R             & \emph{compound(star([atom]))}         & compound(star([atom])) \\
    S             & \emph{atom}                           & atom \\
    \hline\hline
\end{tabular}
\end{table}
%
Afterwards we can update the \prepostspecs{} for \verb|rate_my_ship/2|, but
\verb|ship/1| and \verb|rating/1| are not affected from this. If our program has
no more clauses, the fixed-point is reached, and the analysis stops.

\paragraph{Backtracking.}
\Prespecs{} specify a condition which must be fulfilled at the moment of the call,
and \postspecs{} can provide information about the type of the used terms after
a successful execution.
The caller of a predicate is unaware which clause provided the result.
Thus, the union of all gained type information has to be considered in the second phase.
As a result, it is safe to ignore backtracking:
yet, precision could in some cases be improved if clause ordering
and cuts (\verb|!|) were considered.

\section{Evaluation}%
\label{sec:evaluation}

To our knowledge,
papers on type systems for Prolog usually omit an evaluation of their applicability
for existing, real-world Prolog code
and offer insights on their type inference mechanisms
on small toy examples, such as the well-known \verb|append| predicate.
However, we want to consider code that is more involved than homework assignments.
There is no indication to what extent type inference approaches are applicable to the real world,
or how much work has to be spent re-writing code for full-fledged type systems.

In contrast, we baptise \ourtool{} by fire
and evaluate for how many variables
in the code we can infer a type that is more precise than \verb|any|.
For this, we use smaller SWI community packages\footnote{\url{http://www.swi-prolog.org/pack/list}},
as well as \prob{}~\cite{DBLP:conf/fm/LeuschelB03},
a model checker and constraint solver that currently consists of more than \num{120000} lines of Prolog code.

\subsection{Known Limitations}

Currently, we face three limitations in \ourtool{}:
firstly, as we try to avoid widening whenever possible,
i.e., we try to use the most precise type like a \verb|one_of| instead of generalising to their common supertype,
performance is not too good.
Analysis of small projects runs neglectably fast,
yet \prob{} requires several hours to complete a full analysis.
Secondly, libraries throw a wrench into our scheme:
modern Prolog systems pre-compile the code.
Hence, meta-programs, such as term expanders,
cannot access their clauses.
Thus, library code is not considered
and \ourtool{} has to rely on annotations.
Currently, we only provide annotations
for large parts of the lists library
(for both \swi{} and \sic{})
and the AVL tree library (for \sic{} only).
Otherwise, for all library predicates that are not annotated,
an \verb|any| type has to be assumed.
Thirdly, we currently do not consider disjunctions
and if-then-else constructs,
but may gain additional precision
once this is implemented.

Additionally, there is an inherent limitation in our analysis strategy:
some predicates may really work on \emph{any} type,
e.g. term type checking predicates (such as \verb|ground/1| or \verb|nonvar/1|)
or the \verb|member/2| predicate regarding the first argument.
As no similar analysis for Prolog programs exists yet
and type inference by hand is infeasible for large programs,
it is certainly hard to gauge the precision of our type inference.

\subsection{Empirical Evaluation}

\begin{table}[t]
\caption{Amount of Inferred Types for Variables}
    \label{tbl:results}
    \centering
\setlength{\tabcolsep}{4.5pt}
\begin{tabular}{lrrr}
    \hline\hline
    Repository                   & \# Variables   & Inferred Types    & Unknown Calls          \\
    \hline
    bddem                        & 196            & 31.63 \%          & 57.6 \% \\ 
    dia                          & 400            & 68.5 \%           & 8.23 \% \\
    maybe                        & 32             & 6.25 \%           & 70.0 \% \\
    plsmf                        & 67             & 37.31 \%          & 37.5 \% \\
    quickcheck                   & 122            & 42.6 \%           & 34.1 \% \\
    thousands                    & 19             & 94.73 \%          & 0.0 \% \\
    $\varnothing$ SWI Community Packages        & 68344          & 21.8   \%         & 39.0 \%                    \\
    \prob{}                      & 81893          & 21.2   \%         & 20.8 \%           \\
 \hline\hline
\end{tabular}
\end{table}

    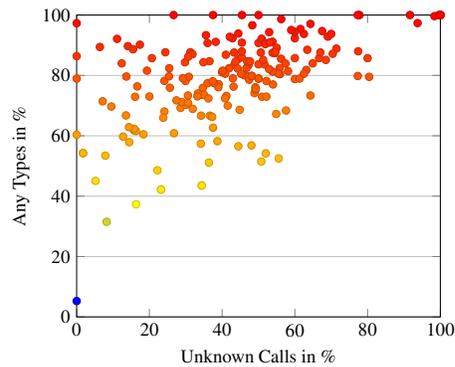
\begin{wrapfigure}[12]{R}{18em}\vspace{-5.5em}
    \centering
    \resizebox{0.5\textwidth}{!}{
\begin{tikzpicture}
\begin{axis}[
   xlabel={Unknown Calls in \%},
   ylabel={Any Types in \%},
   scatter,
   only marks,
   ymajorgrids,
   ymin=0,
   ymax=100,
   xmin=0,
   xmax=100,
   scaled y ticks = false
]

\addplot table[x=unknown, y=any] {data.csv};
\end{axis}
\end{tikzpicture}}
        \vspace{-7pt}
\caption{Correlation Between Unknown Calls and Inferred Types}%
\label{fig:corr}
\end{wrapfigure}

In \cref{tbl:results},
the results of some repositories\footnote{Full results: \url{https://github.com/pkoerner/plstatic-results/tree/wflp-20}}
and the mean value of the 198 smallest community packages is shown.
We give the amount of Prolog variables,
and the percentage of which we can infer a type that
is a strict sub-type of \verb|any|.
For reference, we also give the amount of calls to unknown predicates
in order to give an idea how many missing types
are caused by, e.g., library predicates lacking annotations.
Though, once a variable is assigned
an \verb|any| type,
the missing precision typically is passed on to terms
that are interacting with the \verb|any| term
as the predicate is implemented in a library.

At first glance, the fraction of inferred types
seems to be rather low.
For some repositories,
such as \enquote{dia} and \enquote{thousands},
a specific type could be inferred for a large percentage of variables.
Note that in return, the amount of unknown calls is relatively low.
Then, there are repositories such as \enquote{bddem} and \enquote{plsmf},
which both are wrappers of a C library.
As such, the interop predicates are unknown
and the inferred types are significantly lower.
Finally, there are packages like
\enquote{maybe}, \enquote{quickcheck} and projects such as \prob{},
that make use of other libraries,
conditional compilation, meta-calls and other features
that decrease accuracy of type inference.

Overall, we were surprised how small the amount of inferred types was.
Though, one has to consider
that a large amount of predicates are library calls,
e.g. into the popular CLP and CHR libraries.
In~\cref{fig:corr}, we show this relation.
One can clearly recognise that (unknown) library calls
negatively impact the results of our type analysis.
Yet, many auxiliary predicates are written to be polymorphic
and deal with any type.

With \ourtool{}, we were able to find several errors:
many SWI libraries have been broken
with changes introduced in \swi{} 7~\cite{wielemaker2014swi}.
Strings now are proper strings,
where legacy code relies on the assumption that they
are represented as code lists.
Furthermore, \ourtool{} located calls in \prob{}
that were guaranteed to fail every time due to type errors.
These calls decide whether a backend is usable
in order to solve a given predicate and always fail.
Thus, the errors have gone unnoticed for eight years,
as the backend simply was not used.
One error was reported due to missing term expansion
as we did not execute untrusted Prolog code.
We found another false-positive due to meta predicate annotations
which add the module to a goal, thus altering the term structure.
Additionally, we found some extensions \sic{}
made to the ISO standard that we were not aware of:
e.g., arithmetic expressions in \sic{} allow expressions such as \verb|X is integer(3.14)| or
\verb|Y is log(2, 42)|.
Thus, \ourtool{} raised type errors
for terms that did not match our type describing ISO arithmetic expressions.


%
%
%
%
%
%
%
%
%
%

\section{Conclusion and Future Work}%
\label{conclusion}

In this paper, we presented \ourtool{}, a tool that re-uses its annotations
in order to verify types statically where possible.
%
In several existing Prolog repositories,
\ourtool{} was able to locate type errors.
Yet, without annotations of further libraries,
the amount of actual inferred types remains relatively low.
We invite the Prolog community to discuss
whether such type annotations are desired
and should be shipped as part of packages.

There remains some work on \ourtool{}:
performance bottlenecks need to be reviewed.
Furthermore, the analysis would heavily benefit
from a mechanism for the term expander
to hook into library packages,
manual annotations or generated annotations
based on library source code as far as it is available.
It might also be possible to analyse some pre-compiled library
beforehand and re-use those results in the
analysis of the main program.
We also plan to implement semantics for new types,
for which the structure is not specified, but they
may only be created by libraries.
E.g., Prolog streams
cannot be created manually and
one of the built-in predicates \emph{must} be called.
Other examples include ordered sets or AVL trees,
where it is possible to create or manipulate such a term,
but it is heavily discouraged as it is very easy to introduce subtle errors.

Moreover, it would be exciting to compare
the amount of inferred types to similar implementations
such as CiaoPP.
We assume their analysis to be stronger,
but suspect that Ciao's approach might not scale
as well for larger programs.
Yet, comparison might be hindered, again,
because features of other Prolog systems are not supported.
It might also be interesting to see whether our semantics
can be integrated into CiaoPP.

In~\cite{stulova2016reducing} and also in the evaluation of \plspec{}~\cite{plspec}, it was determined that
the overhead of run-time type checks can be enormous,
especially if applied to recursive predicates.
With additional type information,
a large amount of run-time checks can be eliminated,
as, e.g., proposed by~\cite{stulova2016reducing}.
It is fairly straightforward to
generate a list of already discharged annotations
and use that as a blacklist in \plspec{}.
This could move the tool towards gradual typing~\cite{siek2015refined},
combining benefits of static typing and reducing overhead of static checks
with the potential for many optimisations.

It is well-known that compilers often benefit heavily from
type information.
An interesting research question is to
investigate the impact of type information,
e.g. gained by \ourtool{} or by annotations,
when added to the binding-time analysis of a partial evaluator,
such as \textsc{Logen}~\cite{leuschel2004specialising}.
This might greatly reduce the work required
of manually improving generated annotations
in order to gain additional performance.

As a more pragmatic approach to future work,
it would be greatly appreciated if the state-of-the-art
of Prolog development tooling could be improved.
Currently, IDEs and editor integrations are lacking.
Including type information would be a great start.

\bibliographystyle{abbrv}
\bibliography{paper}

%
%

\end{document}